\documentclass[]{nature}
\makeatletter\if@twocolumn\PassOptionsToPackage{switch}{lineno}\else\fi\makeatother

\usepackage{tabulary,graphicx,amsmath,amsfonts,amssymb}
\usepackage[utf8x]{inputenc}
\usepackage[T1]{fontenc}
\usepackage{siunitx}

\makeatletter
\renewenvironment{figure}
               {\@float{figure}}
               {\end@float}
\renewenvironment{figure*}
               {\@dblfloat{figure}}
               {\end@dblfloat}
\renewenvironment{table}
               {\@float{table}}
               {\end@float}
\renewenvironment{table*}
               {\@dblfloat{table}}
               {\end@dblfloat}

\makeatother

\usepackage{url,multirow,morefloats,floatflt,cancel,tfrupee}
\makeatletter

\AtBeginDocument{\@ifpackageloaded{textcomp}{}{\usepackage{textcomp}}}
\makeatother
\usepackage{colortbl}
\usepackage{xcolor}
\usepackage{pifont}
\usepackage[nointegrals]{wasysym}
\makeatletter

\def\mcWidth#1{\csname TY@F#1\endcsname+\tabcolsep}

\def\cAlignHack{\rightskip\@flushglue\leftskip\@flushglue\parindent\z@\parfillskip\z@skip}
\def\rAlignHack{\rightskip\z@skip\leftskip\@flushglue \parindent\z@\parfillskip\z@skip}

\@ifundefined{etal}{}{}

\usepackage{ifxetex}
\ifxetex\else\if@twocolumn\@ifpackageloaded{stfloats}{}{\usepackage{dblfloatfix}}\fi\fi

\AtBeginDocument{
\expandafter\ifx\csname eqalign\endcsname\relax
\def\eqalign#1{\null\vcenter{\def\\{\cr}\openup\jot\m@th
  \ialign{\strut$\displaystyle{##}$\hfil&$\displaystyle{{}##}$\hfil
      \crcr#1\crcr}}\,}
\fi
}

\AtBeginDocument{%
  \@ifpackageloaded{endfloat}%
   {\renewcommand\efloat@iwrite[1]{\immediate\expandafter\protected@write\csname efloat@post#1\endcsname{}}}{\newif\ifefloat@tables}%
}%

\def\BreakURLText#1{\@tfor\brk@tempa:=#1\do{\brk@tempa\hskip0pt}}
\let\lt=<
\let\gt=>
\def\processVert{\ifmmode|\else\textbar\fi}

\@ifundefined{subparagraph}{
\def\subparagraph{\@startsection{paragraph}{5}{2\parindent}{0ex plus 0.1ex minus 0.1ex}%
{0ex}{\normalfont\small\itshape}}%
}{}

\newcommand\role[1]{\unskip}
\newcommand\aucollab[1]{\unskip}

\@ifundefined{tsGraphicsScaleX}{\gdef\tsGraphicsScaleX{1}}{}
\@ifundefined{tsGraphicsScaleY}{\gdef\tsGraphicsScaleY{.9}}{}
\def\checkGraphicsWidth{\ifdim\Gin@nat@width>\linewidth
	\tsGraphicsScaleX\linewidth\else\Gin@nat@width\fi}

\def\checkGraphicsHeight{\ifdim\Gin@nat@height>.9\textheight
	\tsGraphicsScaleY\textheight\else\Gin@nat@height\fi}

\def\fixFloatSize#1{}
\let\ts@includegraphics\includegraphics

\def\inlinegraphic[#1]#2{{\edef\@tempa{#1}\edef\baseline@shift{\ifx\@tempa\@empty0\else#1\fi}\edef\tempZ{\the\numexpr(\numexpr(\baseline@shift*\f@size/100))}\protect\raisebox{\tempZ pt}{\ts@includegraphics{#2}}}}

\AtBeginDocument{\def\includegraphics{\@ifnextchar[{\ts@includegraphics}{\ts@includegraphics[width=\checkGraphicsWidth,height=\checkGraphicsHeight,keepaspectratio]}}}

\DeclareMathAlphabet{\mathpzc}{OT1}{pzc}{m}{it}

\def\URL#1#2{\@ifundefined{href}{#2}{\href{#1}{#2}}}

\def\UrlOrds{\do\*\do\-\do\~\do\'\do\"\do\-}%
\g@addto@macro{\UrlBreaks}{\UrlOrds}

\edef\fntEncoding{\f@encoding}

\makeatother

\newif\ifmultipleabstract\multipleabstractfalse%
\usepackage[nomarkers,tablesfirst,nolists]{endfloat}

\def\fixFloatSize#1{}
\begin{document}

\author{Qi Shen\textsuperscript{1,2,3,*}, Jian-Yu Guan\textsuperscript{1,2,3,*}, Ji-Gang Ren\textsuperscript{1,2,3,*}, Ting Zeng\textsuperscript{1,2,3}, Lei Hou\textsuperscript{1,2,3}, Min Li\textsuperscript{1,2,3}, Yuan Cao\textsuperscript{1,2,3}, Jin-Jian Han\textsuperscript{1,2,3}, Meng-Zhe Lian\textsuperscript{1,2,3}, Yan-Wei Chen\textsuperscript{1,2,3}, Xin-Xin Peng\textsuperscript{1,2,3}, Shao-Mao Wang\textsuperscript{1,2,3}, Dan-Yang Zhu\textsuperscript{1,2,3}, Xi-Ping Shi\textsuperscript{1,2,3}, Zheng-Guo Wang\textsuperscript{1,2,3}, Ye Li\textsuperscript{1,2,3}, Wei-Yue Liu\textsuperscript{1,2,3,8}, Ge-Sheng Pan\textsuperscript{1,2,3}, Yong Wang\textsuperscript{5}, Zhao-Hui Li\textsuperscript{3,4}, Jin-Cai Wu\textsuperscript{3,4}, Yan-Yan Zhang\textsuperscript{6}, Fa-Xi Chen\textsuperscript{7},  Chao-Yang Lu\textsuperscript{1,2,3}, Sheng-Kai Liao\textsuperscript{1,2,3}, Juan Yin\textsuperscript{1,2,3}, Jian-Jun Jia\textsuperscript{3,4}, Cheng-Zhi Peng\textsuperscript{1,2,3}, Hai-Feng Jiang\textsuperscript{1,2,3}, Qiang Zhang\textsuperscript{1,2,3}, Jian-Wei Pan\textsuperscript{1,2,3}}

\title{113 km Free-Space Time-Frequency Dissemination at the 19th Decimal Instability}

\maketitle

\begin{affiliations}
    \item Hefei National Laboratory for Physical Sciences at the Microscale and School of Physical Sciences, University of Science and Technology of China, Hefei 230026, China
    \item Shanghai Branch, CAS Center for Excellence and Synergetic Innovation Center in Quantum Information and Quantum Physics, University of Science and Technology of China, Shanghai 201315, China
    \item Shanghai Research Center for Quantum Sciences, Shanghai 201315, China
    \item Key Laboratory of Space Active Opto-Electronic Technology, Shanghai Institute of Technical Physics, Chinese Academy of Sciences, Shanghai 200083, China
    \item Xinjiang Astronomical Observatory, Chinese Academy of Sciences, Urumqi 830011, China
    \item Key Laboratory of Time and Frequency Primary Standards, National Time Service Center, Chinese Academy of Sciences, Xi’an 710600, China
    \item Jinan Institute of Quantum Technology, Jinan, Shandong 250101, China
    \item Faculty of Information Science and Engineering, Ningbo University, Ningbo 315211, China
    \\
    $^\ast$These authors contributed equally to this work.
\end{affiliations}

\baselineskip24pt

\begin{abstract}
Optical clock networks play important roles in various fields, such as precise navigation~\cite{mehlstaubler2018atomic, Lisdat2016}, redefinition of “second” unit~\cite{Riehle_2018, riehle2015towards, McGrew2019, Bize2019}, and gravitational tests\cite{kolkowitz2016gravitational}. To establish a global-scale optical clock network, it is essential to disseminate time and frequency with a stability of $10^{-19}$ over a long-distance free-space link. However, such attempts were limited to dozens of kilometers in mirror-folded configuration~\cite{deschenes2016synchronization,sinclair2016synchronization}. Here, we take a crucial step toward future satellite-based time-frequency disseminations. By developing the key technologies, including high-power frequency combs, high-stability and high-efficiency optical transceiver systems, and efficient linear optical sampling, we demonstrate free-space time-frequency dissemination over two independent links with femtosecond time deviation, $3\times10^{-19}$ at 10,000 s residual instability  and $1.6\times10^{-20}\pm 4.3\times10^{-19}$ offset. This level of the stability retains for an increased channel loss up to 89 dB. Our work can not only be directly used in ground-based application, but also firmly laid the groundwork for future satellite time-frequency dissemination.
\end{abstract}

\section*{Main}
The time is one of the seven basic International System of Units. Starting on April 2020, five other basic units have also been relied on the unit of the time/frequency because of the best precision and stability it can achieve~\cite{Riehle_2018}. Indeed, state-of-the-art optical clocks have demonstrated the most accurate frequency reference, up to $10^{-19}$ level~\cite{campbell2017fermi,mcgrew2018atomic} locally. To remotely access the reference optical clocks and establish a global-scale network, it is further necessary to disseminate time-frequency over long distances with an accuracy of a similar level of $10^{-19}$. It has been expected that an optical clock network connected by dissemination technology will open many exciting applications, including the next-generation definition of the unit of “second”, testing general relativity~\cite{kolkowitz2016gravitational, derevianko2014hunting, Delva2017, safronova2018search}, probing the changes of physical constants~\cite{Cheng2009Ultracold}, searching for gravitational waves and dark matter~\cite{kolkowitz2016gravitational, derevianko2014hunting}, and long-range quantum networks~\cite{liu2019experimental}.

So far, optical carrier phase transfer through fiber links with two-way transfer operation has reached over 1,000 km with a stability of $10^{-19}$ at 100~s~\cite{droste2013optical,predehl2012920}. However, such a method is incompatible with the existing telecommunication fibers, and can be difficult to reach certain locations such as mountains\cite{katori2011optical}, marine and intercontinental ranges, and spaces. In particular, to establish a global-scale network, it is essential to develop free-space time-frequency dissemination. On this path, previous work have achieved a distance up to a dozen kilometers \cite{giorgetta2013optical,deschenes2016synchronization,Bodine2020,Shen2021} in a mirror-folded configuration, which cannot meet the high demands required by future satellite-ground time-frequency dissemination.

Compared to the fiber links, the free-space channel in the troposphere region are much more unstable due to various atmospheric disturbance, which can lead to frequent signal dropouts. Therefore, the dissemination system needs a large ambiguous range to enable link reconnection without accumulated error. The continuous-wave (CW) laser carrier used in fiber time-frequency links typically has an ambiguous range of a few fs, which is too small to avoid cycle slips after signal dropouts. The pulsed laser-based ranging method has a sufficiently long ambiguous range~\cite{Samain2015, Exertier2019}, however, amplitude detection with a photodiode or a single-photon detector usually limits the instability of the link to the ps level. To overcome these challenges, here, we develop a number of novel optical frequency comb-based techniques, which combines features of high-precision optical phase detection and a large ambiguous range. Further, we employ linear optical sampling that offers femtosecond accuracy over the whole ambiguous range. By using 1-W optical frequency combs and nW-level linear optical sampling modules, we achieved a time-frequency dissemination over a free-space link of 113 km and obtain a stability better than $3\times 10^{-19}$ at 10,000 s. 

\begin{figure}
\centering
\includegraphics[width=0.9\columnwidth]{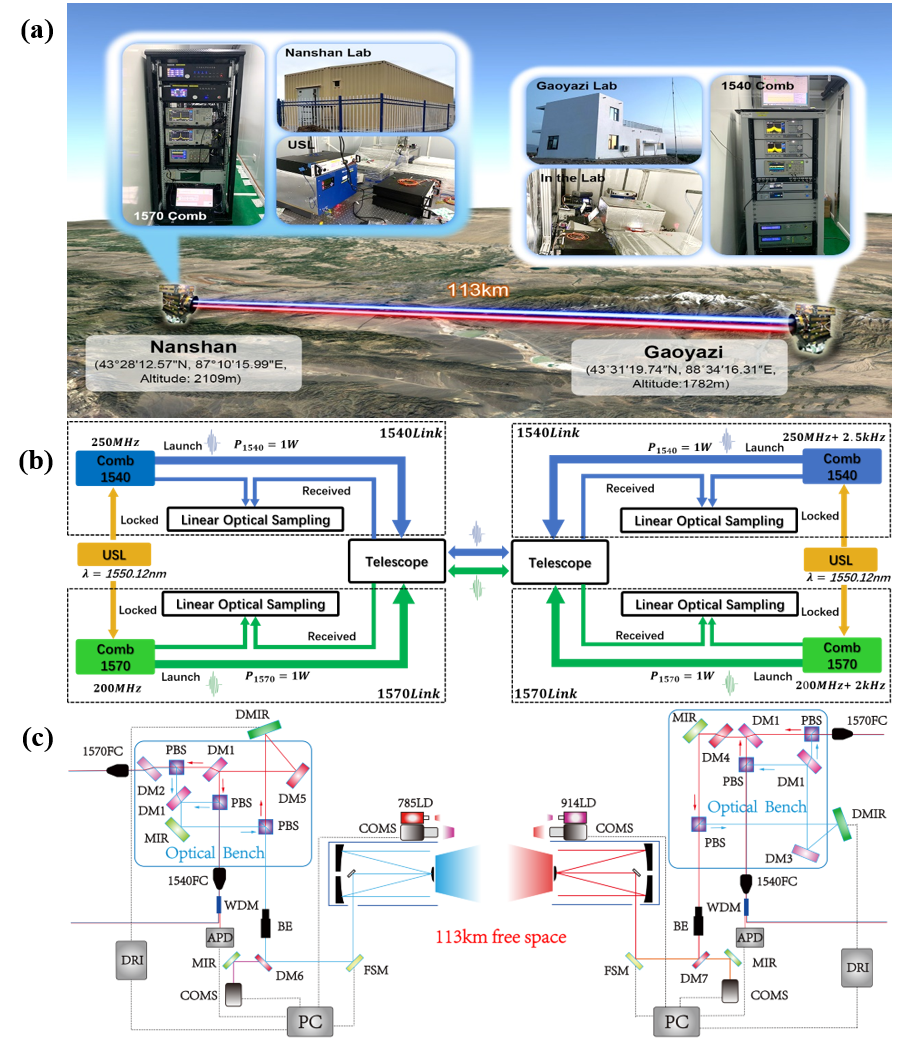}
\caption{The experimental setup. (a). Overview of the 113 km free space time-frequency dissemination experiment. (b). The main diagram of the system, which contains a 1540 nm link and a 1570 nm link. USL: ultra stable laser. (c). The optical layout of the optical transceiver telescope, where different collors of optical paths indicate different polarizations. PBS: polarizing beam splitter; 1540FC (1570FC): fiber collimator for two-way 1540 (1570) nm comb transceiver; DM: dichroic beam splitter; MIR: mirror; DMIR: deformable mirror; WDM: wavelength division multiplexer; BE: beam expander; FSM: fast steering mirror; DRI: controller of DMIR; COMS: complementary metal–oxide semiconductor; PC: processing computer.}
\label{Fig:Setup}
\end{figure}

The experiment was performed from mid-July to mid-August in 2021 in Urumqi city, Xinjiang Province. As shown in Fig.~\ref{Fig:Setup}, two terminals (A and B) are located at Nanshan and Gaoyazi in Urumqi city with a distance of 113 km.  Each terminal is equipped with an ultrastable laser, two high-power optical frequency combs (OFCs) with different wavelengths centered at 1540 nm and 1570 nm, two linear optical sampling (LOS) modules, and an optical transceiver telescope. An ultrastable laser (USL) with a $3\times10^{-15}@  1 s$ frequency instability at the wavelength of 1550.12 nm is used as the reference clock source. The OFC optical phase locked to the USL is used as the carrier and reference signal of local sampling. 
Instead of keeping two terminals close and folding links with a remote flat mirror in the previous experiments \cite{giorgetta2013optical, deschenes2016synchronization, Bodine2020, Shen2021}, we physically separate the two terminals to a distance of 113 km. The folded configuration is convenient for link performance evaluation; however, the time-frequency signal has not actually been disseminated over a long distance physically. Obviously, the results in the previous implementations may benefit from the similar environment and common reference clock and thus can't simulate the real challenging environment for satellite-based free-space time-frequency dissemination.
By frequency multiplexing the common free-spece channel, we establish two independent two-way time-frequency transfer links, enabling precise evaluation of the link performance without limitation from the USL.

The repetition rate of the 1540 nm OFC is 250 MHz, and the differencial repetition between two terminals is set to 2.5 kHz, while the repetition rate of 1570 nm OFC is 200 MHz, and the differencial repetition rate is set to 2 kHz. To overcome the high loss of the link, two-stage high-power EDFAs are exploited, providing a 1 W output after filtering with a 20-nm spectrum bandwidth, which is approximately 5-10 times higher than the OFC power in the previous experiments. The 3-dB bandwidth of both 1540 nm OFC and nm 1570 OFC is approximately 7 nm, determined by the gain region of the EDFA amplifer. Short pulse corresponds to low noise and high signal-to-noise interference; however, high peak power can induce unexpected nonlinearity and damage. Considering that the systems have to continuously opertate over a long time, we produce a chirped pulse to reduce the power density to avoid burns. The pulse widths at the output are between 60-90 ps for the 1540 and 1570nm OFCs (See methods for detail). Water coolers are used to protect the EDFAs from overheating; in addtion, they reduce the temperature fluctuaiton by a factor of approximately 10.

The optical part of the LOS (see Fig.~\ref{Fig:Setup} b) is fully made of pig-tailed optical components housed in an  aluminum box. To minimize the thermal drift effect on the nonreciprocal fiber path, the length of mismatched fibers is optimized to be below 5 cm, and the temperature is stabilized with a thermoelectric cooler (TEC), exhibiting a standard deviation of 7 mK, coresponding to a time deviation below 0.03 fs. The interference data between the local and received OFC signals are detected by using low-noise balanced detectors, digitized with analog-to-digital converters (ADCs), and recorded with a field-programmable gate array (FPGA) to determine the laser pusle arrival time. All electronics share the same reference clock signal of repetition rate of the local comb. Thanks to GPS receivers at both terminals, the deviation of the time synchronization error between two terminals is approximately 30 ns. The LOS can determine the arrival time of the received OFC according to a single overlapping process between two OFCs. Acctually, only if the recorded data exhibit a clear interaction peak can the arrival time be determined precisely. Here, we set a threshold to select the valid data.  80\% valid data can be obtained simultaneously at both terminals when the average received power is approximately 6 nW.

Two dedicated optical transceiver telescopes with automatic direction tracking functions are developed for this experiment. Each telescope in the two terminals is equipped with a beacon laser for direction tracking at wavelengths of 785 nm and 914 nm, respectively. The primary mirror of each telescope is a Cassegrain reflector with an aperture of 400 mm and focal length of 1,600 mm. We employ an orthogonal polarization scheme to separate sending and receiving OFCs, providing precise measurement of the receiving OFC power. An orthogonal polarization setting is designed to provide large isolation of two-way transfer, and it can also simulate the nonreciprocal situation in satellite ground links due to the relative angular motion between a satellite and its ground station. Such a design induces non-reciprocal optical path. To minimize instability induced by the nonreciprocal paths, we bonded all optical components on a plate of fused silica, and the temperature stabilized to be approximately 1 degree peak-to-peak. There is a deformable mirror inserted in the receiving optical path to manipulate the adaptive optics to compensate for incoming wavefront distortions caused by the atmosphere, improving the coupling efficiency from free space into a single mode fiber.

\begin{figure}
\centering
\includegraphics[width=1\columnwidth]{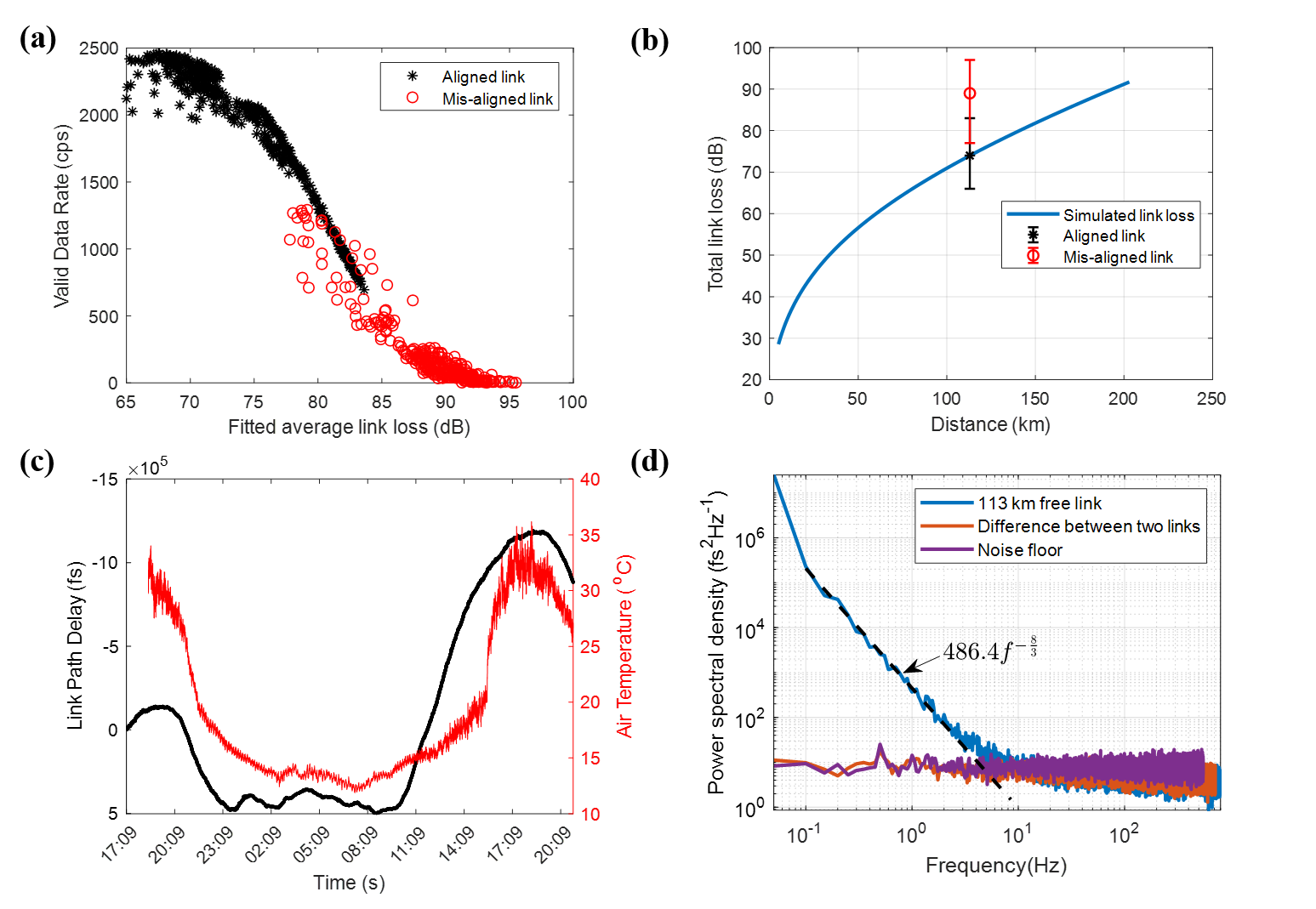}
\caption{Characterization of the 113 km free space link. (a). (Color online) Relation between link loss and valid data rates of the 1540 nm link. (b). Link loss at different distance. (c). Optical length measured by the 1540 nm link and temperature measured at terminal B. (d). Power spectral densities of the free running and compensated link delay. The noise floor is measured by connecting two terminals with a 10 m fiber.}
\label{Fig:Result2}
\end{figure}

The key point of performing the experiment over a long link is to conquer large link loss, so that recording the link loss is important to characterize the link condition. However, the stray reflection of the local OFC is usually much stronger than the recieved OFC; thus, it cannot be measured directly. To monitor the link loss, we develop an approach of determining link loss according to peak voltages of the LOS, valid data rates and statistical loss distribution (see Methods for details). The accuracy of this approach is calibrated by using a commercial optical power meter. Fig.~\ref{Fig:Result2}a shows the relation between the link loss and valid data rates (Sn, amount per sceond). Each point is an avarage value measured in 160 seconds. Typically, loss of the aligned 113 km link fluctuates between 66 dB and 83 dB, with an average value of 74 dB. The maximum link loss of 83 dB corresponds to a valid data rate of more than 600 cps, i.e. about 25\% of the total.  To evaluate the loss tolarence, we deliberately mis-aligned the optical path to reduce the average loss up to 89 dB. Accually, even when the average loss exceeds 90 dB, corresponding to a 1 nW avarage received power, valid data rates can be more than 100 cps. The highest valid data rates, close to 100\% for aligned link and up to 50\% for mis-aligned one, always occur at twilight, while the lowest data rates appear at noon due to the strong atmospheric vortex caused by sunshine heating. The distribution of the link loss is different between the two working conditions, while the dynamic ranges are almost the same. The link loss is also estimated theoretically according to our telescope and atmospheric parameters, as shown in Fig.~\ref{Fig:Result2}b (see Methods for details), and it is largely consistent with the measured link loss. According to the simulation, the 89 dB average loss of the  mis-aligned link  corresponds to that of a 190 km link.

The link delay flucatuation fundamentally limits the link instability because the link delay results limit noise cancellation, and it can be obtained from timing data. Fig.~\ref{Fig:Result2}c shows that the link delay drifts more than 1 ns peak-to-peak over 100,000 s, (about 28 hours) attributed to temperature fluctuation. We choose a 200 s link path delay data of Fig.~\ref{Fig:Result2}c to further analyze atmospheric noise in the frequency domain, resulting in the noise power spectra density (PSD)(see Fig.~\ref{Fig:Result2}d). At low frequencies below 3 Hz, the PSD of the free link exhibits a single power-law decay at $f^{-8/3}$, where $f$ is the Fourier frequency, which is dominated by the turbulence-induced atmospheric piston effect and consistent with the proposed atmospheric model\cite{Strohbehn1978, Conan:95, giorgetta2013optical}. Considering the two-way operation of the link, the transfer function from free link noise to residual link noise is similar to a one-order low pass filter with a bandwidth of light speed over link distance. Thus, the turbulence-induced time jitter is about 10 atto-seconds, which is negligible. As a result, the link noise is the same as the noise floor measured by a shorted-fiber link with the two terminals located in a common laboratory, as shown in Fig.~\ref{Fig:Result2}d. The noise floor is  attributed to downconverted high frequency noise related to the laser pulse width, sampling rates and  dispersion management.

\begin{figure}
\centering
\includegraphics[width=1\columnwidth]{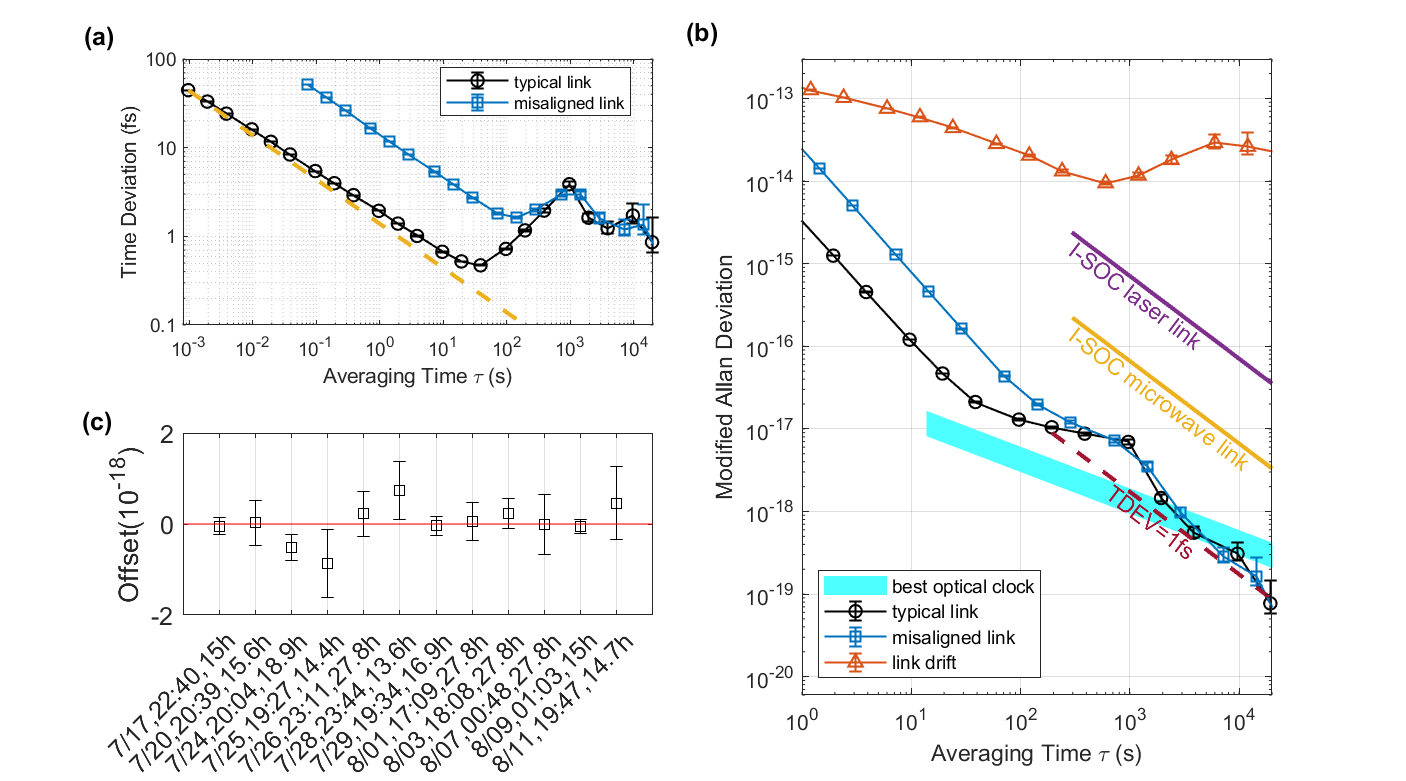}
\caption{Experimental results of time-frequency transfer. (a). (Color online) The time deviation (TDEV) of one link.  (b) The fractional frequency instability of links. Black circles show instability of the well-aligned 113-km free-space time-frequency link; blue squires  show instability of the mis-aligned link; orange triangles show instability of the free running link. Note that we suppose that two independent links have similar performance, and the instability of one link is divided by $\sqrt{2}$. The performance of the best optical clock~\cite{schioppo2017ultrastable,oelker2019demonstration}, I-SOC laser link, I-SOC microwave link and TDEV of 1 fs are also shown. Here, the original data of the best optical clocks is the Allan deviation (ADEV). We suppose the modified Allan deviation (MDEV) is $\sqrt{2}$ times smaller provided that the noise type is white frequency noise~\cite{calosso2016avoiding}. (c) The fractional offset, or bias, across 12 datasets, each label led by start time and duration. The uncertainty per point is estimated as the MDEV value at 10,000 s for the same data set. The red line is the weighted average.}
\label{Fig:Result1}
\end{figure}

Having fully characterized the 113 km link, we carried out a time-frequency transfer experiment over approximately 3 weeks. We obtained a total of 12 data sets, and the continuous acquisition duration of each data set was between 13.6 and 27.8 hours. The link can operate most of the time, except during the rainy period or at noon on sunny days. The instability of the link is evaluated by comparing the two independent links, and we assume that the two links have the same performance. Fig.~\ref{Fig:Result1} a shows the time deviation (TDEV) of the link. It averages down with a slope of $\sqrt{\tau}$ depending on the amount of valid data for short terms and reaches 0.5 fs at 40 s. The precision of valid data does not exhibit an obvious correlation to the received optical power; the start points of both TDEV curves are around 40 fs. For long terms, the thermal drift effect dominates TDEV to be around 1 fs, and bumps at around 1000 seconds correspond to the control cycle of air conditioners. Note that the thermal drift effect can be further improved by passively matching the optical path, thermal stability or active stabilization of the optical phase.

The frequency transfer performance is presented by the modified Allan deviation (MDEV), as shown in Fig.~\ref{Fig:Result1}b. The relative instability is below $3\times10^{-19}$ at 10,000 s and in the range of $10^{-20}$ at 20,000 s. Even for the mis-alinged link, the instability reaches $10^{-19}$ at a few thousand seconds.  Compared with the best atomic clock, the link is more stable for integration time longer than one hour.  The link instability is about 40/$\sqrt{\tau \times Sn}$ determined by random noise of the laser system and sampling process for short terms; while it is a few $10^{-15}/\tau$ limited by the thermal effect for long terms. The instability is more than 3 orders of magnitude better than that of commonly used microwave links \cite{Bauch2006, Fujieda2018} and more than one order of magnitude better than that of underdeveloped multichannal microwave links and laser links \cite{Cacciapuoti2017}. 

The systematic fractional offset is another important parameter for evaluating the performance of the time-frequency transfer link in addition to the stability, and it can be obtained by calculating the nonzero slope to the measured link residual time offsets. The fractional offset shown in Fig.~\ref{Fig:Result1}c is below  $4.3\times10^{-19}$ (1$\sigma$) with a weighted average of $1.6\times10^{-20}$, according to 12 times experimental results. The performance of such a link meets the requirements of an upcoming new SI metrology system. A mendatory step of SI second revision is to compare different atomic clocks with instability below $5\times10^{-18}$, which can be done in one hour.

In summary, we have implemented a time-frequency comparison over a 113 km free space link. The TDEV of the link is around 1 fs for long terms, and the relative instability is below $3\times10^{-19}$ at 10,000 s, while the link loss up to 89 dB. Several key techniques, particularly Watt level OFC, the separation of receiving and transmitting light using an orthogonal polarization scheme, and high sensitivity linear optical sampling detection, have been employed and verified towards satellite ground time frequency dissemination. Based on these techniques, we anticipate that long-haul free space OFC links, combined with fiber-based and satellite-based time-freqeuency links, will become important parts of future optical clock networks. 

\section*{Methods}
\begin{figure}[p]
\centering
\includegraphics[width=0.9\columnwidth]{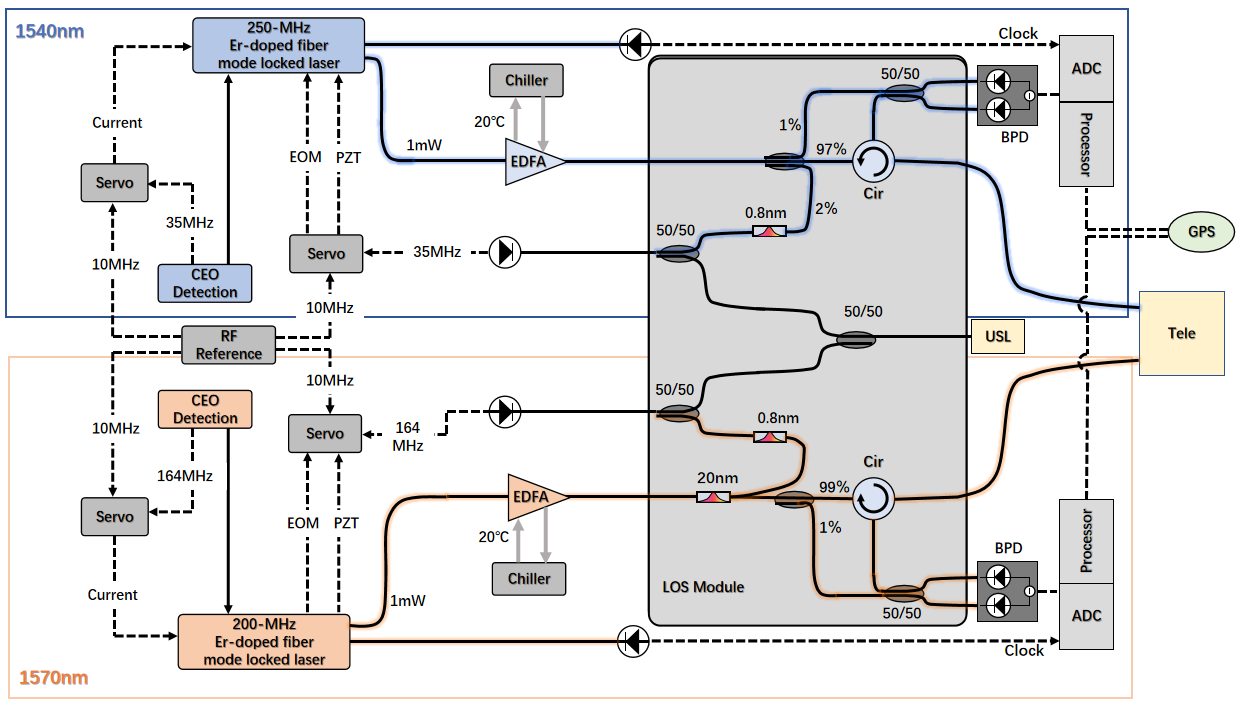}
\caption{Detailed experimental optics setup of single terminal. USL, ultra-stable laser; EDFA, erbium-doped fiber amplifiers; Cir, circulator; BPD, balanced photodiode; Tele, Telescope.}
\label{Fig:OpticalSetup}
\end{figure}

\textbf{Detailed experimental optics setup and parameters} 
The experimental setup of one terminal is shown in Fig.~\ref{Fig:OpticalSetup}. The setup of two terminals are identical except the repetition rate of the optical frequency combs (OFCs). There are two OFCs and one ultra-stable laser (USL) at each terminal. The USL worked as the reference clock and the two OFCs are phase-locked to the it. The upper part of the figure is the OFC bought from Menlo System. Its wavelength is in $1540\pm10$~nm and we call it 1540 OFC for convenience. The bottom one is the homemade OFC. Its wavelength is in $1570\pm10$~nm and we call it 1570 OFC for convenience. For both OFCs, a high-power EDFA is used to amplify the power to about 1~Watt. The detail about the apmlifiers can be seen in the last section. We built an small integrated optical fiber module to achieve the optical beat between the OFCs and the USL, as well as the linear optical sampling (LOS). For 1540 OFC, a 1X3 beam splitter (BS) divide the beam into three parts. The 1\% part is used as local beam of the LOS. The 97\% part is connected to the telescope. The last 2\% part goes through a 0.8~nm filter and then beat with the USL. For 1570 OFC, it first goes through a 20~nm filter centered at 1570~nm. The reflect part further passes a 0.8~nm filter and then beat with the USL. The passed light is then divided by a 1/99 BS. The 1\% part is used as local beam of the LOS and the 99\% part is connected to the telescope. All non-reciprocal optical path lies in the integrated module. Its temperature is stabilized at a standard deviation about 10 mK.

We used two independent analog-to-digital converters (ADCs) and field programmable gate array (FPGA) to record and process the signals. The reference clock of these electronics is come from the repetition rate of corresponding OFC. A GPS receiver provides time synchronization service so that all  the electronics at two terminals can share a same starting time. This is crucial for synchronize the aquisition data at two terminals. In each terminal, a rubidium radio frequency (RF) reference synchronize all the locking electronics. The carrier-envelope offset (CEO) frequency and the beat frequency is set equal for each OFC. The purpose is stated as below.

When all phase-locked loop closed, we have the relation of $f_{cw}+f_{beat}=f_{ceo}+N*f_{rep}$, where N is an integar. Here, we set $f_{ceo}$ as same as $f_{beat}$, to make $f_{cw}=N*f_{rep}$. All microwave frequencies are canceled so that the repetition rate of the OFC is only depends on the frequency of the USL. Frequencies are listed in Table.~\ref{Table:Parameters}. We define repetition rate of 1540nm OFC at terminal A (Nanshan) is 250 MHz. The frequency of ultra-stable laser and the repetition of Comb 1570 at terminal A can be determined by phase-locking parameters. Actually, all frequencies are dependent on frequency of the ultra-stable laser, and daily drift between two ultra-stable lasers is several kilohertz.

\clearpage

\clearpage
\begin{table}[p]
\centering
\resizebox{\textwidth}{!}{%
\begin{tabular}{ll}
\hline
Frequency of the ultra-stable laser at terminal A, $f_{cw,A}$                               & $192.401 THz$           \\
Repetition rate of Comb 1540 at terminal A, $fr_{1540,A}$                                   & $250 MHz$               \\
Index of the tooth of Comb 1540 locking to the ultra-stable laser at terminal A, $N_{40,A}$ & 773604                  \\
Repetition rate of Comb 1570 at terminal A, $fr_{1570,A}$                                    & $199.99504 MHz$         \\
Index of the tooth of Comb 1570 locking to the ultra-stable laser at terminal A, $N_{70,A}$ & 967029                  \\ \hline
Reference frequency of the ultra-stable laser at terminal B, $f_{cw,B}$                     & $f_{cw,A}-1.8675 MHz$   \\
Repetition rate of Comb 1540 at terminal B, $fr_{1540,B}$                                   & $fr_{1540,A}+2582.9 Hz$ \\
Index of the tooth of Comb 1540 locking to the ultra-stable laser at terminal B, $N_{40,B}$ & 773596                  \\
Repetition rate of Comb 1570 at terminal B, $fr_{1570,B}$                                   & $fr_{1570,A}+2066.2 Hz$ \\
Index of the tooth of Comb 1570 locking to the ultra-stable laser at terminal B, $N_{70,B}$ & 967019                  \\ \hline
Difference of repetition rate of the two Comb 1540, $\Delta fr_{1540}$                      & $2582.9 Hz$             \\
Difference of repetition rate of the two Comb 1570, $\Delta fr_{1570}$                      & $2066.2 Hz$             \\
Real frequency of the ultra-stable laser at terminal B, $f_{cw,real,B}$ & $[f_{cw,B}-5.38 kHz,f_{cw,B}-2.39 kHz]$ \\
\hline
\end{tabular}%
}
\caption{The detailed parameters of the experiment.}
\label{Table:Parameters}
\end{table}

\clearpage

\textbf{The link loss determination}
In this experiment, the ratio between sent and received OFC power is up to a few $10^9$, three orders higher than the isolation between sending and receiving optical path (approximately 60 dB). Thus, the received power can not be measured simultaneously while the link works. Here, we develop an approach of acquiring the received OFC power according to the peak voltage of the LOS.

\begin{figure}[p]
\centering
\includegraphics[width=0.9\columnwidth]{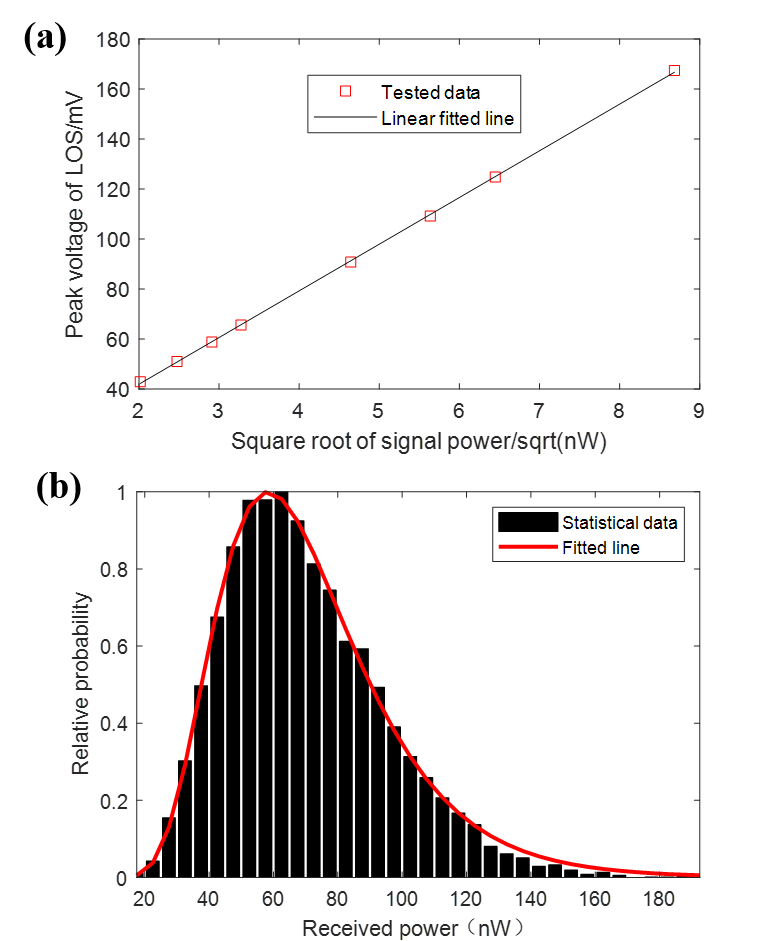}
\caption{(a). The relationship of peak voltage of LOS and the square root of received signal power. (b). The measured signal power distribution and the fit to the data using log-normal curve.}
\label{Fig:PeakV}
\end{figure}

The relationship between the peak voltage and received OFC power is shown in Fig.~\ref{Fig:PeakV}(a). There is good linear relation between the peak voltage and square root of the received OFC power, i.e., $V_{peak}+V_{offset}\propto \sqrt{P_s}$. Based on this relationship, we can inversely calculate the power. However, it is not straight forward, because the LOS record signals only if the peak values exceed the threshold of around 40 mV, corresponding to a power of about 4 nW. We have to estimate the avarage power according to the valid data ratio and loss distrubition, while the link loss is high.

The distribution of the link loss is measured at dusk, so that we have nearly 100\% valid data to fit it. As shown in Fig.~\ref{Fig:PeakV}(b), probalility of the received OFC well fits a log-normal distribution,consistent with atmospheric model given by \cite{Milonni_2004}. Thus, we can estimate the average power based on this distritbution and the valid date ratio.

\begin{figure}[p]
\centering
\includegraphics[width=0.9\columnwidth]{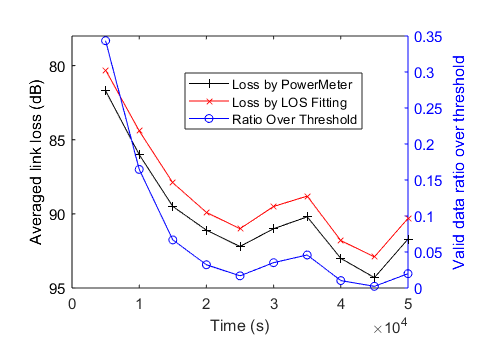}
\caption{The average link loss calculated by the both the power meter monitoring method and the peak voltage fitting method.}
\label{Fig:LinkLossComp}
\end{figure}

In order to evaluate the accuarcy of this approach, we calibrate it with a commercial power meter (S154C of Thorlabs Inc.). An optical coupler is inserted at the 1540 nm LOS of terminal A to guide 5\% power to a power meter, the sending OFC power is reduced to avoid stray reflection effect, and the local OFC power at the LOS is kept same. We perform the measurement 10 times, and each one lasts 5,000 seconds. Fig.~\ref{Fig:LinkLossComp} gives the link loss obtianed with two methods. The peak fiting approach gives a slightly lower value in comparision with that given by the power-meter; and the difference is below 2 dB. The corresponding valid data ratio is shown here as well.



\clearpage

\clearpage

\clearpage

\textbf{Link loss simulation of the horizontal free space channel}
The loss of the whole link from input of LOS of one terminal to input of the photodetector of the other terminal can be expressed by Eq.~\ref{link loss}.

\begin{align}\label{link loss}
\eta_{link}&=\eta_{los}\eta_{tele\_link}\notag\\
&=(\eta_{los\_t}\eta_{los\_r})(\eta_{tele\_t}\eta_{geo}T_{atm} \eta_{tele\_r} \eta_{sm}\eta_{atp})
\end{align}

The link loss mainly contains two parts, the LOS optics loss $\eta_{los}$ and the telescope link loss $\eta_{tele\_link}$. The LOS optics loss, including the transmitter loss $\eta_{los\_t}$ and receiver loss $\eta_{los\_t}$, are around 4 dB. The telescope link loss includes other parts. $\eta_{tele\_t}$ and $\eta_{tele\_r}$ are the telescope optical efficiency of the transmitter and receiver, respectively.

The geometric attenuation is expressed as

\begin{equation}
  \eta_{geo}=(\frac{D_r}{D_r+L \theta_{t}})^2,
\end{equation}

where $D_{r}$ is the apertures of the receiver telescope; L is the free space channel distance; $\theta_{t}$ represents the effective transmitter full-angle divergence for the free space link.

$T_{atm}$ is the atmospheric transmittance\cite{Strohbehn1978}, which is reduced by the air absorption and scattering of the propagating beam, described by the expression,

\begin{equation}
  T_{atm}=e^{-\sigma L},
\end{equation}

where $\sigma$ is total extinction coefficient containing the absorption and scattering coefficients of atmospheric molecules. Since the NIR band is called the 'atmospheric window', atmospheric molecular absorption is not considered in non-rainy and foggy weather. Atmospheric molecular scattering can be handled by Rayleigh scattering theory\cite{Kim2001}.
 
\begin{equation}
  \sigma=\frac{3.91}{V}(\frac{\lambda}{550nm})^{-q}
\end{equation}

V is visibility (km), $\lambda$ is the optical wavelength, $q$ is the size distribution of the scattering particles. 

\begin{equation}
\begin{cases}
	q=0.58{V}^{\frac{1}{3}}, & \text{${V} \le 6km $}\\[2ex]
	q=1.3,  & \text{$ 6km<{V}<50km $} \\[2ex]
	q=1.6,  & \text{${V} \ge 50km $} \\[2ex]
\end{cases}
\end{equation} 

$\eta_{sm}$ represents the single-mode fiber coupling efficiency. It is related to the parameter $r_0$, also called Fried’s parameter\cite{Dikmelik2005}, is usually defined by the expression,

\begin{equation}
  r_0=[0.42sec(\zeta)k^2\int_{h_0}^{H}C_n^2dh]^{-3/5}
\end{equation}

where $\zeta$ is zenith angle, $k$ is wave number of beam wave, $C_n^2$ is refractive-index structure parameter, and $H$ is the altitude of the higher site, which equals to $h_0+Lcos(\zeta)$, $h_0$ represents the altitude of the lower site.

Theoretical formula of calculating the free-space single-mode coupling efficiency\cite{Andrews2005}.

\begin{equation}
  \eta_{sm}=8a^2\int_{0}^{1}\int_{0}^{1}\exp[-(a^2+\frac{A_R}{A_C})(x_1^2+x_2^2) \times I_0(2\frac{A_R}{A_C}x_1x_2)x_1x_2dx_1dx_2]\\
\end{equation}

where

\begin{equation}
  a=\frac{D_r}{2}\frac{\pi{W_m}}{\lambda{f}}
\end{equation}

\begin{equation}
  A_R=\frac{\pi{D_r^2}}{4}
  \end{equation}

\begin{equation}
  A_C=\pi\rho_c^2
  \end{equation}
 
 \begin{equation}
  \rho_c^2=(1.46C_n^2k^2L)^{-3/5}
  \end{equation}
  
a is the ratio of the receiver radius to the radius of the backpropagated fiber mode. $D_r$ is the receiver lens diameter, $W_m$ is the fiber-mode field radius at the fiber end face, $\lambda$ is the optical wavelength and $f$ is the focal length of the receiver. $A_R$ is the area of the receiver aperture,  $A_C$ is the spatial coherence area of the incident plane wave, also called its speckle size. $\rho_c$ is the spatial coherence distance.

$\eta_{atp}$ is the decrease in efficiency arising from the ATP and atmospheric turbulence, which is typically 3 dB.

\begin{figure}[p]
\centering
\includegraphics[width=0.9\columnwidth]{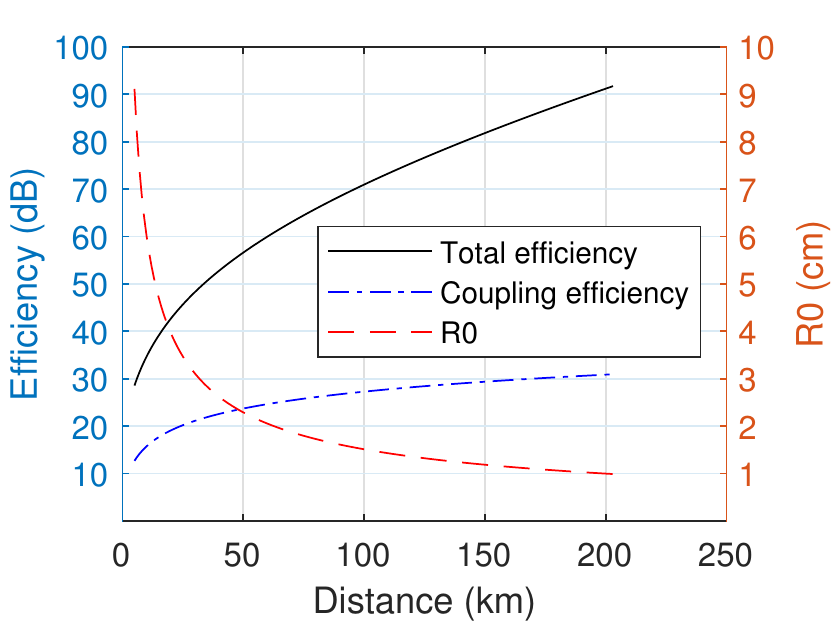}
\caption{Extended Data Figure: Simulated r0, single-mode coupling efficiency and total attenuation change with distance.}
\label{Fig:LinkCalc}
\end{figure}

Based on our system setup and the 113 km free space link, the input parameters for the total link calculation is shown in Table.~\ref{tab1}. The telecope link loss is estimated to be 70 dB, including the telescope optical efficiency of 4.8 dB, the geometric attenuation of 21.8 dB, the atmospheric attenuation of ~12.5 dB, the single mode coupling attenuation of 27.9 dB, and the ATP (Aqcuisition, Tracking and Pointing) system related loss of 3 dB. Thus, the total link loss is calulated to be 74 dB. The simulated r0, single-mode coupling efficiency and total link loss is shown in Fig.~\ref{Fig:LinkCalc} with distance changes. In a typical nice weather, the link loss is measured to fluctuate between 66 dB and 83 dB, with the average of 74 dB, which is consistent to the calculation value.

\clearpage
\begin{table}[p]
\caption{Input parameters for the total link calculation}
\label{tab1}
\centering
\begin{tabular}{|l|l|l|}
\hline
$\theta_{t}$	&	divergence angle 	& 40 $\mu$rad \\\hline
V						&	visibility 					& 40 km   \\\hline
$\lambda$		& wavelength				& 1550 nm  \\\hline
$C_n^2$  		& refractive-index structure& 4E-16\\\hline
$D_r$				& receiver lens diameter		& 400 mm  \\\hline
$f$				& focal length of the receiver		& 1.584 m  \\\hline
$W_m$		& fiber-mode field radius		& 5 $\mu$m \\ \hline
\end{tabular}
\end{table}

\clearpage

The total link loss of the 113 km horizontal free space channel does not reach the tolerated maximum value of our system. According to the simulation results, the total telescope link loss of 190 km distance is 89 dB, which is comparable with the average telescope link loss of the mis-aligned link curve in Fig.2 of the main text. So it can be predicted that the limit transmission distance of our system should be around 200 km.

\clearpage

\begin{figure}[p]
\centering
\includegraphics[width=0.9\columnwidth]{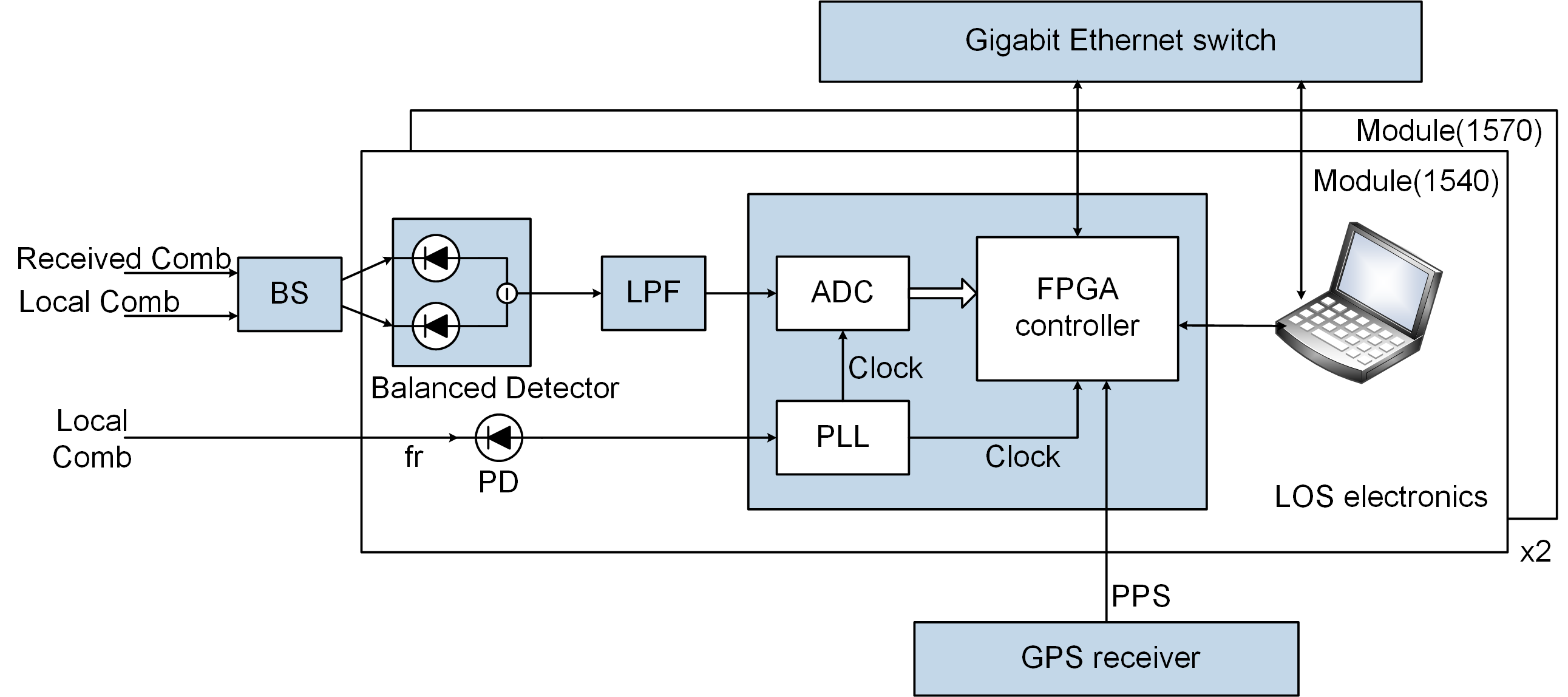}
\caption{Architecture of the LOS electronic system.}
\label{Fig:LosElec}
\end{figure}

\textbf{Linear sampling electronics system}
The architecture of the LOS electronic system is illustrated in Fig.~\ref{Fig:LosElec}, consisting of a Global Position System (GPS) receiver, a Gigabit Ethernet switch and two sets of LOS electronic module, one for 1540 comb system and another for 1570 comb system. Inside the LOS electronic module, the received comb pulses and local comb pulses interfere at a 50:50 optical beam splitter (BS), and a balanced detector with 150MHz bandwidth is used to convert the optical signal into an electric signal. After low-pass filtered, the interferogram voltages are digitized by a 14 bit analog-to-digital Converter (ADC), and then the digitized data stream is transmitted to a Field Programmable Gate Array (FPGA), which is a Zynq-7000 XC7Z045 here, responsible for the kernel digital processing. Both the ADC sampling rate and the working clock of FPGA are synchronized with the repetition frequency of the local comb pulse, via a photoelectric detector (PD) and phase-locked loop (PLL) circuits integrated in the hardware.  

The GPS receiver is applied to synchronize the start time between different sites, by generating a pulse per second (PPS) to the LOS electronic module with 30 nanosecond time precision. Once the FPGA controller receives the PPS, a 64-bit binary counter starts to work. When the amplitude of the interferogram signal is above a given threshold, current timestamp is recorded, and through peak searching function implemented in the FPGA, a data window of 1024 sampling points around the interferogram peak is extracted and transferred to the computer via gigabit Ethernet for data storage and analysis. 

As mentioned above, one data frame is 1024 sampling points, and each point is a 16-bit binary data, assumed that the repetition frequency difference between the two combs is 2.5kHz, a data rate of 5M byte per second will be generated. In this case, it will take a lot of time to process the whole data after hours of data storage, therefore real time data processing is utilized. Once an hour of data is stored, the phase calculation program implemented in the MATLAB will run, then the processed result which mainly contains the timestamp and the measured arrival time of each site will be transferred to the central server for further processing. Subtraction of the two arrival times yields the clock time offset.

\clearpage

\clearpage

\begin{figure}[p]
\centering
\includegraphics[width=0.9\columnwidth]{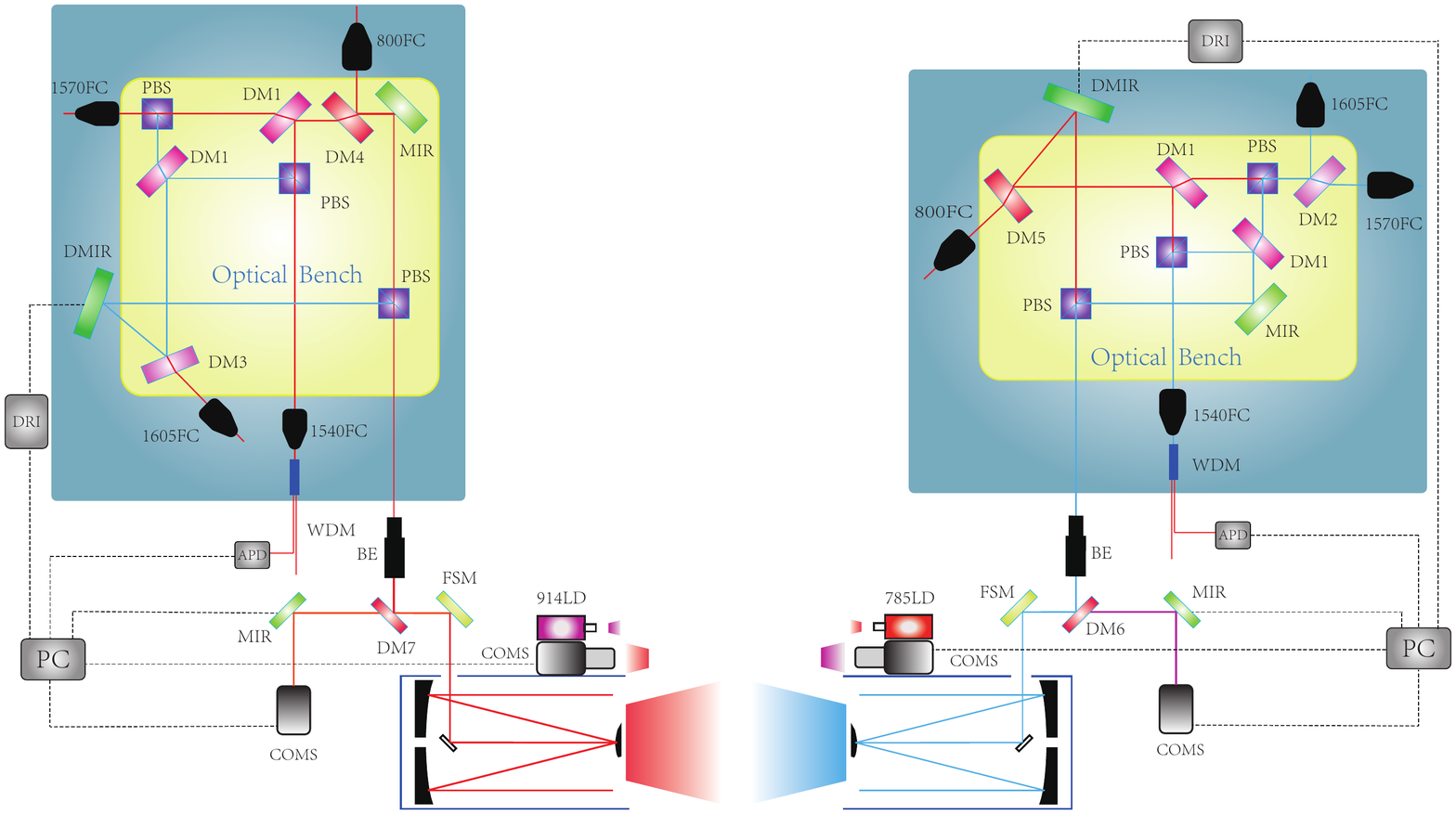}
\caption{The optical transceiver telescope.}
\label{Fig:Telescope}
\end{figure}

\textbf{Optical transceiver telescope}
Our experiment is to simultaneously use two optical combs with central wavelengths of 1540 nm and 1570 nm, separately, for optical two-way time and frequency transfer over free space, which requires the establishment of two independent atmospheric links. A schematic layout of the transceivers that integrated the optical components, electronics and the telescope is shown in Fig.~\ref{Fig:Telescope}. The scheme satisfies the demand of 100 km space precision time–frequency transmission, while verifying the impact of the asymmetric path between the transmitter and receiver on the transmission accuracy due to forward-targeting at the ground transceivers of the satellite-ground time-frequency transmission. We use three polarizing beam splitters (PBS) to design asymmetric transmit and receive paths, preparing the light source with a single horizontal or vertical polarization. In one transceivers terminal, the H-polarized beam passes through the transmit path, the V-polarized beam passes through the receive path, and vice versa in the other terminal. The fixed-focus collimator 1540FC (1570FC), the 1540 nm (1570 nm) optical comb and the devices between them are connected by polarization-maintaining fiber to ensure high H(V) polarization contrast. Reflector (MIR) in the transmit path simulates a ground-based forward mirror for forward targeting.

The path difference between the transmitting and receiving paths is affected by the change in flight time due to temperature fluctuations, and our experimental scheme use optical bench to build asymmetric transmitter and receiver to solves this problem. The materials of optical devices such as PBS, MIR and dichroic beam splitter (DM) are fused silica with low thermal expansion coefficient ($5\times10^{-7}/^\circ C$). Yellow rectangle on the Fig.~\ref{Fig:Telescope} are large fused silica platforms, one size 170*150*30 mm and the other size 190*140*30 mm. In addition, asymmetric transmit and receive are integrated to minimum, ensuring that the asymmetric path is less than 150 mm. The deformable mirror (DMIR), which integrates electronics and optics, is not on fused silica platform, because devices of such high complexity are difficult to integrate with optical bench. We added temperature control in the blue rectangular part (Fig.~\ref{Fig:Telescope}) to reduce the optical paths variations of both the deformation mirror and the optical bench. 

DMIR is added to the receiver to build adaptive optics (AO) system that is beneficial to improve the receive efficiency of single-mode fibres by correcting the distortion of beam induced by atmospheric turbulence. The DRI in Fig.~\ref{Fig:Telescope} is controller of DMIR. Wavelength Division Multiplexing (WDM) separates the beacon light in the 1540 nm signal, and then the beacon light is converted into an electrical signal by avalanche photo-detector (APD), which is further used as the input signal for the AO system. DM6 and DM7 reflect 70\%-75\% of the beacon light into the receiver, and pass 25\%-30\% to the complementary metal–oxide semiconductor (CMOS). A fast steering mirror (FSM), DM6(DM7), MIR and CMOS are used to create a fine tracking system, which dynamically adjusts the optical path according to a correction program with the input of image information obtained by the fine tracking CMOS. Another camera and telescope console to build a coarse tracking system. The beacon light center wavelengths is 785 nm and 915 nm, respectively. The primary mirror of the transceiver terminal is a reflecting Cassegrain telescope with an aperture of 400 mm and $\times$16 magnification. Telescope magnification is not enough, so a $\times$3 magnification beam expander (BE) is added in the transceiver, the total magnification is $\times$48. In addition, we also designed 800FC and 1605FC to transmit and receive optical signals with central wavelengths of 800 nm and 1605 nm for laser communication.

\clearpage

\clearpage

\begin{figure}[p]
	\centering
	\includegraphics[width=0.9\columnwidth]{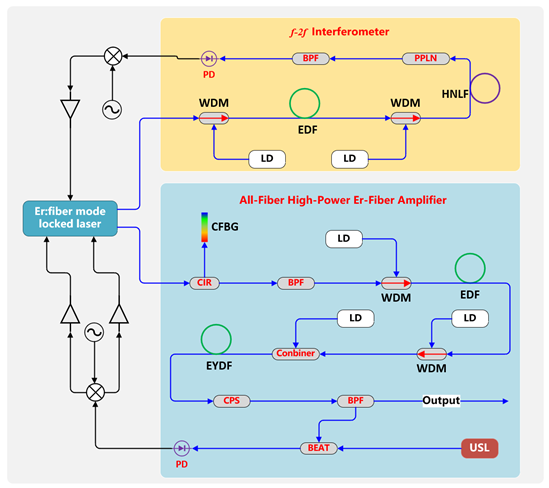}
	\caption{Schematic of high-power frequency comb. WDM, wavelength division multiplexer; ISO, isolator; EDF, Er-doped fiber; LD, laser diode; HNLF, PM-highly nonlinear fiber; PPLN, periodically-poled lithium niobite crystal; BPF, band pass filter; PD, photo detector; CIR, circulator; CFBG, Chirped fiber Bragg grating; Combiner, pump and signal combiner; CPS, cladding power stripper; USL, ultra-stable laser.}
	\label{Fig:comb}
\end{figure}

\textbf{High-power low phase noise optical comb}
The high precision time-frequency link measurements presented here involve four high-power frequency combs based on Er:fiber mode locked lasers. These frequency combs are almost identical with the exception of the repetitions rate and central wavelengths. Therefore, we only introduce 1570-nm comb. A schematic of the frequency combs used is shown in Fig.~\ref{Fig:comb}. The key elements of frequency combs are fiber oscillator, carrier-envelope offset frequency detection and the pulse amplifier. The oscillator was mode-locked with an all polarization-maintaining nonlinear amplifying loop mirror. The PZT and EOM were inserted in the laser cavity for repetition rate control. The average output power of the oscillator was $\sim$1 mW and the pulses allowed for sub-100 fs pulse duration if compressed in standard fiber.  

A part of the pulse train of the oscillator ($\sim$0.5 mW) is used for determine the carrier-envelope offset frequency of the Er:fiber oscillator. The oscillator was followed by an EDFA, a 15-cm HNLF for octave-spanning supercontinuum generating, and a PPLN to double the supercontinuum light at 2000 nm. The common-path f-2f interferometer was adopted for carrier-envelope offset frequency measurement. The signal-to-noise ratio of carrier-envelope offset frequency signal is more than 35 dB in a 300 kHz resolution bandwidth (RBW).

The pulse amplifier consisted of the pulse stretcher, the preamplifier and the power amplifier. The pulse stretcher adopted Chirped fiber Bragg grating with a reflectivity of 50\% and a bandwidth of 60 nm centered at 1560 nm, resulting in a pulse duration of 120 ps. And then the light is launched into a 20-nm band pass filter. To compensate for the loss of Chirped fiber Bragg grating and the 20-nm band pass filter, the pulse train was amplified in a preamplifier, that is forward and backward pulped with 800 mW of 976-nm laser diodes. The pulses are amplified up to 120 mW. The power amplifier comprised of a 200-cm long polarization maintaining double-clad Er-Yb co-doped fiber with a 10-$\mu$m core diameter and $130-\mu$m cladding diameter that was forward-pumped by a high-power fiber-coupled multimode diode at 976 nm through a signal-pump combiner. The cladding power stripper (CPS) is placed after the power amplifier to strip the residual pump. The fiber of output port of CPS and the second 20-nm band pass filter are PM1550, which is single mode at 1550 nm. 

\begin{figure}[p]
	\centering
	\includegraphics[width=0.9\columnwidth]{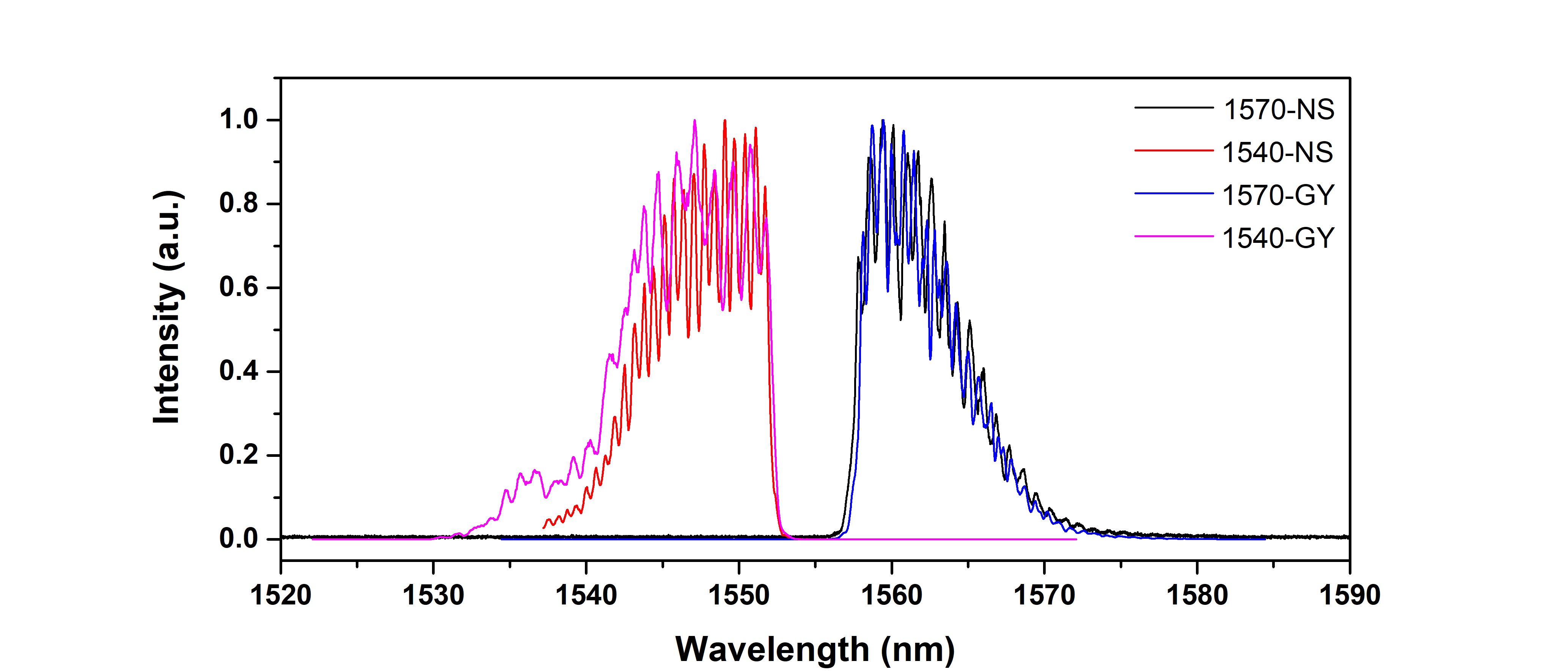}
	\caption{Measured spectrum of high-power amplifiers.}
	\label{Fig:spectrum}
\end{figure}

\begin{figure}[p]
	\centering
	\includegraphics[width=0.9\columnwidth]{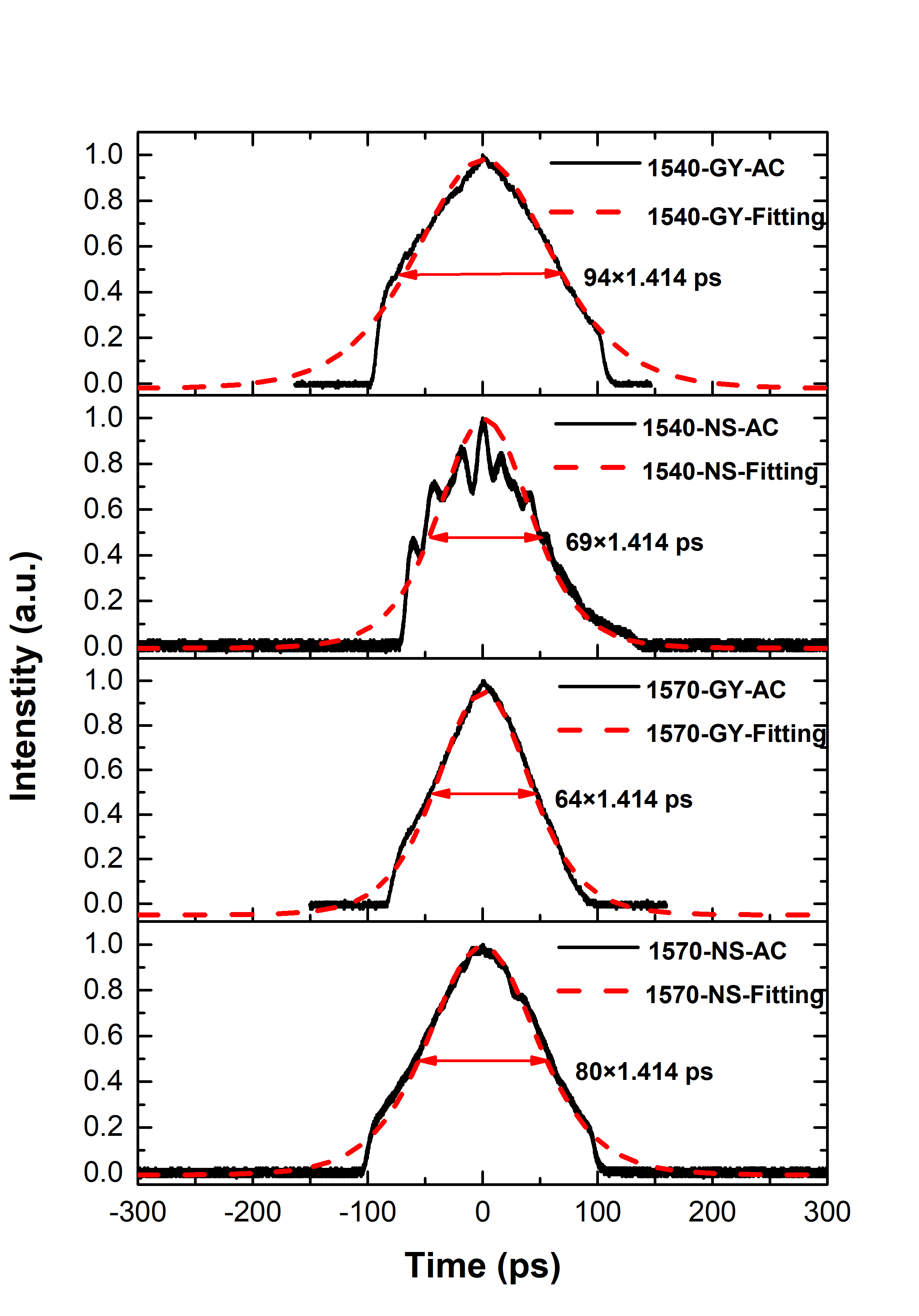}
	\caption{Autocorrelation trace of high-power amplifiers.}
	\label{Fig:AC}
\end{figure}

\begin{figure}[p]
	\centering
	\includegraphics[width=0.9\columnwidth]{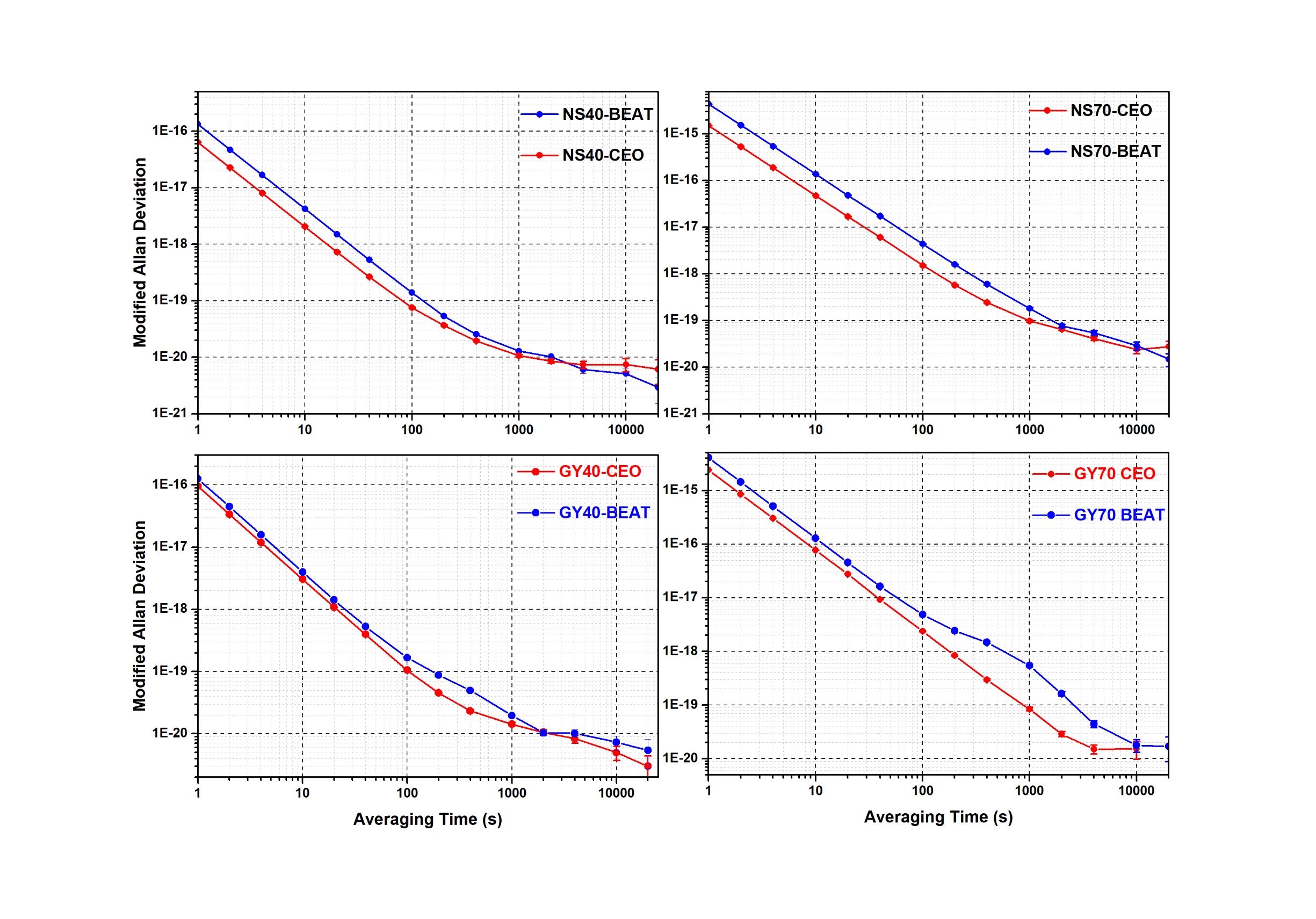}
	\caption{Fractional frequency stabilities of combs.}
	\label{Fig:NS}
\end{figure}

After amplification, we achieved more than 1W of an average output power. The amplified pulses had a 20 nm FWHM(full width half maximun) bandwidth centered at 1540 nm or 1570 nm and durations of 60-100 ps assuming as gaussian-form profile. Figure~\ref{Fig:spectrum} shows the measured spectra of the pulses from BPF. The measured autocorrelation traces and the corresponding Gaussian fitting curves of the power amplifier is shown in Fig.~\ref{Fig:AC}. In order to evaluate the characterization of comb stabilization, the carrier envelop frequency and beat frequency are measured by directly counter at 1 second gate time. The fractional frequency stability is plotted in Fig.~\ref{Fig:NS} . 

\section{Data availability}
The data that support the plots within this paper and other findings of this study are available from the corresponding author upon reasonable request.

\section{Code availability}
All relevant codes or algorithms are available from the corresponding author upon reasonable request.

\section*{References}

\bibliographystyle{naturemag}

\bibliography{FT_Reference}

\section*{Acknowledgements}
This work is supported by the National Key Research and Development Program of China (grant no. 2017YFA0303900, 2020YFA0309800, 2020YFC2200103); Strategic Priority Research Program of Chinese Academy of Sciences (grant no. XDB35030000, XDA15020400);  National Natural Science Foundation of China(grant no. T2125010, 61825505); Anhui Initiative in Quantum Information Technologies (grant no. AHY010100); Key Research and Development Program of Guangdong Province (grant no. 2018B030325001); Shanghai Municipal Science and Technology Major Project (grant no. 2019SHZDZX01).

\section*{Authors' contributions}
All authors contributed extensively to the work presented in this paper.

\section*{Competing interests}
The authors declare no competing interests.

\section{Authors' information}

\end{document}